\documentstyle[twocolumn,aps,epsfig]{revtex}

\begin{document}
\draft
\preprint{December 11, 2000}
\title{Numerical Replica Limit for the Density Correlation of the Random Dirac Fermion}
\author{Shinsei Ryu$^{1}$ and Yasuhiro Hatsugai$^{1,2}$}
\address{
$^1$Department of Applied Physics, University of Tokyo,
7-3-1 Hongo Bunkyo-ku, Tokyo 113-8656, Japan\\
and\\
$^2$PRESTO, JST, Saitama 332-0012, Japan}

\maketitle

\begin{abstract} 
The zero mode wave function of a massless Dirac fermion 
in the presence of a random gauge field is studied.
The density correlation function is calculated numerically 
and found to exhibit power law in the weak randomness
with the disorder dependent exponent.
It deviates from the power law and the disorder dependence becomes frozen
in the strong randomness.
A classical statistical system is employed through the replica trick to interpret the results
and the direct evaluation of the replica limit is demonstrated numerically.
The analytic expression of the correlation function and the free energy 
are also discussed with the replica symmetry breaking
and the Liouville field theory.
\end{abstract}

\narrowtext

Although the scaling theory of localization for two-dimensional disordered systems 
generally predicts the absence of extended states,
we have some examples of non-localized states in two dimensions which are marginally allowed to appear.
Among them, systems with chiral symmetry such as the Anderson bond-disordered model \cite{gade},
the random flux model \cite{sugiyama,avishai,furusaki}
or the $\pi$-flux model with link disorders \cite{hatsugai,morita}
have attracted lots of attention.
Density of states of the models have singularities at the zero energy
and the corresponding wave functions exhibit multifractal behavior.
Actually, one generally expects the existence of zero energy states for these models.
Consider a Hamiltonian of the above models with chiral symmetry
${\cal H}=\sum_{<ij>}c_{i}^{\dagger}t_{ij}c_{j} + h.c.$
on a bipartite lattice $\Lambda$ 
which can be decomposed into two sublattices $\Lambda_{A}$ and $\Lambda_{B}$.
After performing a unitary transformation (redefinition of indices), 
the Hamiltonian is expressed as 
\begin{eqnarray}
{\cal H}=
\pmatrix{
\{c_{A}^{\dagger}\} & \{c_{B}^{\dagger}\} \cr
}
\pmatrix{
{\bf 0}_{A} & {\cal D}_{AB} \cr
{\cal D}_{AB}^{\dagger} &{\bf 0}_{B} \cr
}
\pmatrix{
\{c_{A}\} \cr
\{c_{B}\} \cr
}.
\nonumber
\end{eqnarray}
The off-diagonal block structure of ${\cal H}$ implements the fact that hopping is restricted 
between the interpenetrating sublattices $\Lambda_{A}$ and $\Lambda_{B}$.
The zero mode wave functions $\psi={}^{t}(\psi_{A},\psi_{B})$ 
satisfy the Schr$\ddot{o}$dinger equations
\begin{eqnarray}
\label{ldirac}
{\cal D}_{AB}\psi_{B}=0,\hspace{1cm}{\cal D}^{\dagger}_{AB}\psi_{A}=0.
\end{eqnarray}
Let $N_{A,B}$ is the number of sites on $\Lambda_{A,B}$.
For the cases where $N_{A}-N_{B} > 0$
( which can be realized, for example, by an appropriate boundary condition ),
standard linear algebra tells us there always exist $(N_{A}-N_{B})$ 
independent zero energy solutions for Eq. (\ref{ldirac})
with vanishing $\psi_{B}$.
In this expression, the notion of chiral symmetry is explicit.
However, one needs to solve Eq. (\ref{ldirac}) numerically to go further.

\begin{figure}
\begin{center}
\epsfig{file=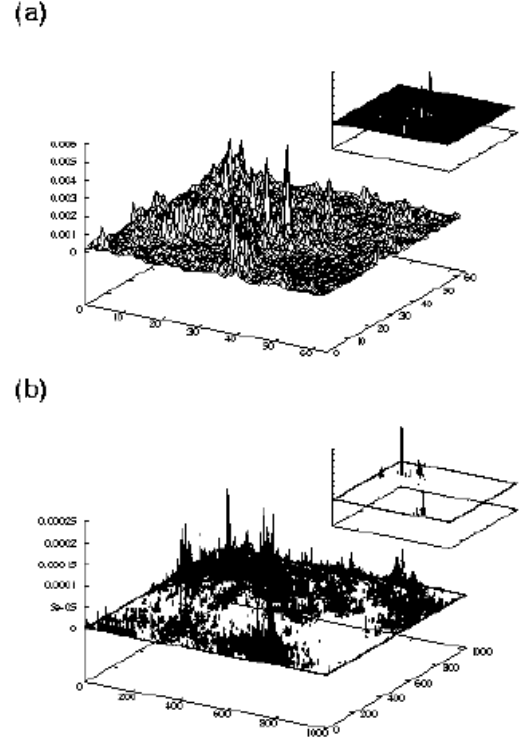,width=8.0cm}
\caption{\label{den}
Typical probability densities $|\psi(x)|^{2}$ for $g=0.4$ on a
(a) $64 \times 64$ lattice and (b) $1024 \times 1024$ lattice.
Insets: $|\psi(x)|^{2}$ for $g=6.4$ for each system.}
\end{center}
\end{figure}

Now, let us concentrate our attention on cases
where the low-lying physics is described by the Dirac fermions.
Remarkably, in this situation, 
the explicit construction of zero energy wave functions is possible 
in a two-dimensional continuum space 
\cite{ludwig} \cite{aharonov}.
Let us consider a Hamiltonian of the form 
$H={\bf \sigma}\cdot{\bf p}+{\bf \sigma}\cdot{\bf A} $
where $\sigma_{i=x,y}$ are the usual $2\times 2$ Pauli matrices and ${\bf A}$ is a random gauge field.
The Schr$\ddot{o}$dinger (Dirac) equations for the zero modes are
\begin{eqnarray}
\label{cdirac}
( 2\partial+i\bar{A} )\psi_{-}=0,\hspace{1cm} ( 2\bar{\partial}+iA )\psi_{+}=0
\nonumber
\end{eqnarray}
where $\partial\equiv( \partial_{x}-i\partial_{y})/2$,
$\bar{\partial}\equiv( \partial_{x}+i\partial_{y} )/2$,
$A\equiv A_{x}+iA_{y}$
and $\bar{A}\equiv A_{x}-iA_{y}$.
This is a continuum analogue of Eq. (\ref{ldirac}).
If we adopt the Coulomb gauge to express the vector potential ${\bf A}$ 
in terms of a scalar potential $\Phi$ as
$A_{x}=\partial_{y}\Phi$, $A_{y}=-\partial_{x}\Phi$
and assume that the mean total flux  piercing the system is zero, 
we can obtain the exact solution for any realization of disorders as
$
\psi_{\pm}({\bf x})=C_{\pm}e^{\mp\Phi({\bf x})}
$.
Further, let us assume the probability weight for each realization 
of $\Phi({\bf x})$ has the form
$
P[\Phi] \propto  e^{-S}
$
where
$
S[\Phi]=1/2g\int d^{2}x (\nabla \Phi ({\bf x}))^{2}
$
and $g$ is the disorder strength or, in the field theoretic language, 
the coupling constant, which is {\em dimensionless } 
in two dimensions.

From now on, we concentrate our attention only on $\psi_{+}$.
For physical interest, 
it is necessary to consider normalized wave functions in a $L \times L$ box as
$
\psi({\bf x})
=e^{-\Phi({\bf x})}/ \sqrt{Z}
$
with
$
Z= \int \frac{ d^{2}{\bf x} }{ a^{2} } e^{-2\Phi ({\bf x}) }
$
where $a$ is a lattice constant.
Here, we regularized the problem on a $N\times N$ periodic lattice
although the original problem is formulated in a continuum space
(where $L=Na$)
.
Correspondingly, we use the probability weight
$
S[\Phi]\propto  
1/2g\sum_{<ij>}(\Phi_{i}-\Phi_{j})^{2} 
$
or, in the momentum space,
$
S[\Phi]\propto  
N^{2}/g\sum_{m}
\widetilde{\Phi}_{m} \widetilde{\Phi}_{-m}
\left( 2-\sum_{\mu =1}^{2}\cos(a k_{m}^{\mu} ) \right) 
$
where 
$
\widetilde{\Phi}_{m}\equiv
N^{-2}\sum_{j}e^{-ik_{m}x_{j}}\Phi_{j}
$
and the sum extends over the first Brillouin zone
$m^{\mu}=-N/2+1\cdots N/2$
with
$
k_{m}^{\mu}=2\pi m^{\mu}/aN
$
($N$ is even for convenience.)\cite{rem0}.
Typical probability densities $|\psi(x)|^{2}$ calculated numerically are shown 
in Fig. \ref{den},
which remind us of multifractal states 
found at a localization-delocalization transition for several systems
\cite{huckestein}.

In fact, 
the multifractal property of this wave function 
has been revealed quantitatively  
by a close analogy to a generalized random energy model\cite{wen,castillo}.
As the disorder strength $g$ varied,
the multifractal spectrum exhibits a sharp transition 
which is similar to the freezing phenomenon in spin glasses.
Several other approaches
such as the supersymmetry (SUSY) technique \cite{mudry}
, the connection to the Liouville field theory \cite{tsvelik}
, the renormalization group (RG)\cite{carpentier}
or conformal field theory \cite{gurarie}
have also been taken to support the transition.

Since the calculated probability densities (Fig. \ref{den}) are so spiky,
the discretization procedure above may not be justified.
In spite of this subtlety, 
we concentrate on this well defined discretized wave function
to investigate universal properties.

In this letter,
we evaluate the density correlation function
\begin{eqnarray}
\label{correl}
\left< \psi^{2}(1)\psi^{2}(2) \right> 
=\left<
\frac{1}{Z^{2}}
 e^{-2\Phi({\bf x}_{1})} e^{-2\Phi({\bf x}_{2})}
\right>
\nonumber
\end{eqnarray}
where $\left<\cdots \right>$ denotes the averaging 
with respect to the weight $P[\Phi]$.
Here, the difficulties reside in the normalization factor $Z$ in the denominator 
since $Z$ itself is a random variable.
The one of the simplest attempts to cope with it is the replica trick.
We multiply the numerator by $Z^{n}$
and consider
\begin{eqnarray}
\label{rep}
\left< \psi^{2}(1)\psi^{2}(2) \right>_{n}
\equiv
\nonumber \\
\int \frac{d^{2}\xi_{1}}{a^{2}}\cdots \frac{d^{2}\xi_{n-2} }{a^{2}}
\left< 
e^{
-2\left[
\Phi({\bf x}_{1} )+\Phi({\bf x}_{2})
+\Phi({\bf \xi}_{1})+ \cdots +\Phi({\bf \xi}_{n-2})
\right]
}
\right> 
\end{eqnarray}
which is expected to reduce to 
$
\left< \psi^{2}(1)\psi^{2}(2) \right> 
$
by taking the replica limit $n\rightarrow0$ (analytic continuation).
We use this replica trick to interpret the direct numerical results
and also try to take the replica limit 
by evaluating 
$\left< \psi^{2}(1)\psi^{2}(2) \right>_{n}$
numerically for several $n$ and extrapolating them to $n=0$.
In addition, 
we utilize
the evaluation of $\left< \psi^{2}(1)\psi^{2}(2) \right>_{n}$
together with the Liouville field theory
to get the analytic expression of the correlation function 
for the weak disorder regime.

\begin{figure}[t]
\begin{center}
\epsfig{file=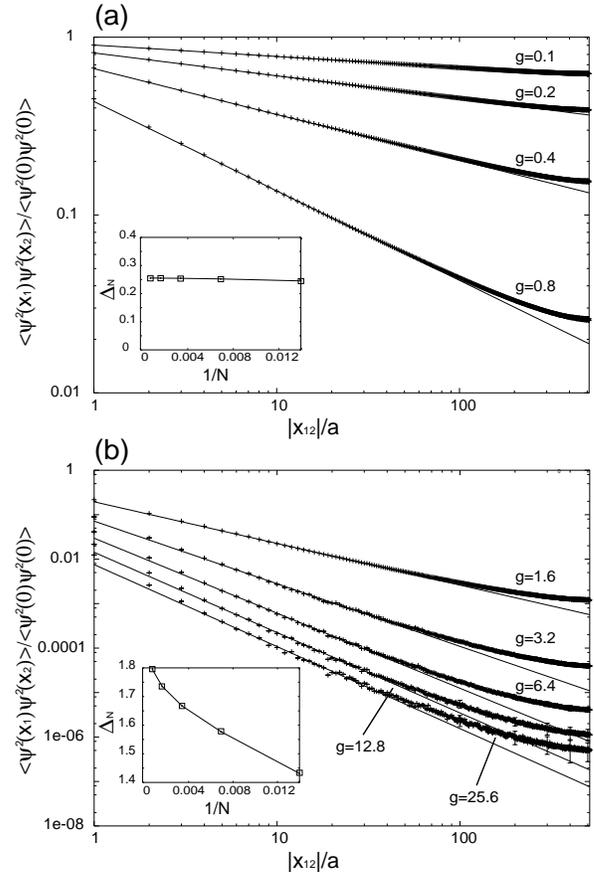,width=8.0cm}
\caption{\label{corr} 
The numerically calculated density correlation for
(a) the weak disorder regime and (b) the strong disorder regime
on a $1024 \times 1024$ lattice.
For $g\le 1.6$, the statistical error is smaller than the data point.
For $g\ge 3.2$, the error bars are shown for every 100 points.
Insets: The ``exponent'' $\Delta_{N}$ v.s. $1/N$ for 
(a) $g=0.4$ and (b) $g=6.4$.
It is evaluated for $1\le |x_{12}|/a \le 10$ for each finite system.
}
\end{center}
\end{figure}

\begin{figure}[t]
\begin{center}
\epsfig{file=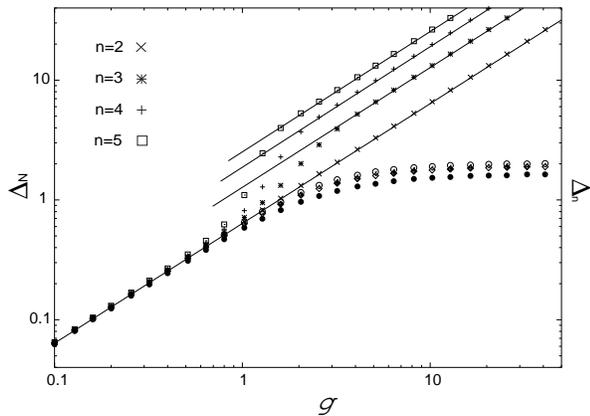,width=8.0cm}
\caption{\label{exp} 
The ``exponent'' $\Delta_{N}$ of the two-point correlation function 
with respect to the disorder strength $g$ 
and its naive replica estimates $\Delta_{n=2,3,4,5}$ (log-log plot).
The ``exponent'' is evaluated for $1\le |x_{12}|/a \le 10$ 
for $64 \times 64$ ($\circ$), $256 \times 256$ ($\Diamond$)
and $1024 \times 1024$ ($\bullet$) system.
The statistical error is smaller than the symbol size.
The analytic estimation Eq. (\ref{delta}) for $\Delta_{n}$ is represented by lines.}
\end{center}
\end{figure}

First, let us present the direct numerical calculation 
of $\left<\psi^{2}(1)\psi^{2}(2) \right>$.
In this problem, 
the probability weight itself is diagonal in the momentum space (see above), 
which allows us to carry out numerical simulations
with a very large lattice 
up to $2048\times 2048$. 
Fig. \ref{corr} shows calculated
$\left<\psi^{2}(1)\psi^{2}(2) \right>$
for various $g$ on a $1024\times 1024$ lattice.
The quenched averaging is performed over $\sim 10^{5}$ different
realization of disorders \cite{rem}.
As is shown, the correlation function for the weak disorder
exhibits power law behavior 
$
\left<\psi^{2}(1)\psi^{2}(2) \right>\sim |x_{12}|^{-\Delta}
$
for $1\lesssim |x_{12}|/a \ll N/2$
with its exponent $\Delta$ dependent linearly on $g$, $\Delta= 2g/\pi$ (Fig. \ref{exp}).
It is consistent with the several analytic approaches \cite{mudry}.
As $g$ increases, however,
the $g$ dependence of the correlation function becomes weaker
and it deviates from the power law.
To be more precise, 
there is a systematic deviation from the simple power law, that is,
if we determine the ``exponent'' $\Delta_{N}$ on a finite $N$ system,
it seems to diverge as $N$ increases ( see the insets of Fig. 2 ).
It is clearly different from the behavior of $\Delta_{N}$ in the weak randomness 
where $\Delta_{N}$ seems to converge.
In fact, as is shown in Fig. \ref{den}, 
the wave function becomes peaked on few sites as $g$ increases.
However, it is different from the usual localized wave function 
which decays exponentially with its typical length scale
characterized by the localization length.
The above change of behavior in the correlation function 
is consistent with the transition from the weak to strong disorder 
found in the multifractal spectrum by the previous studies.

\begin{figure}[t]
\begin{center}
\epsfig{file=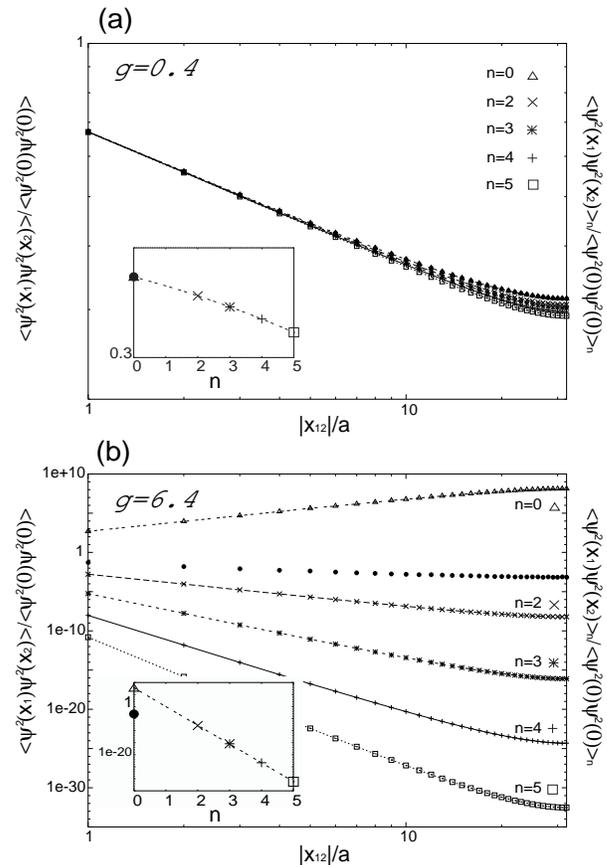,width=8.0cm}
\caption{\label{replica} 
The numerically calculated density correlation ($\bullet$)
and their replica estimates with the different number of replicas 
$n=2,3,4,5$
for (a) $g=0.4$ and (b) $g=6.4$ 
on a $64\times 64$ periodic lattice (log-log plot).
The replica estimates are obtained by 
evaluating the multiple integral in Eq. (\ref{stat}) directly.
$\left<\psi^{2}(x_{1})\psi^{2}(x_{2})\right>_{n=0}$ is obtained 
by extrapolating from $\left<\psi^{2}(x_{1})\psi^{2}(x_{2})\right>_{n=2,3,4,5}$.
Insets: 
$\left<\psi^{2}(x_{1})\psi^{2}(x_{2})\right>_{n}$ v.s. $n$ 
at $|x_{12}|/a=20$ (semi-log plot).
}
\end{center}
\end{figure}

Next, let us try to interpret the above numerical results by the replica trick.
After replicating $Z$, we can perform the averaging $\left< \cdots \right>$ 
in Eq. (\ref{rep})
and obtain
\begin{eqnarray}
\label{stat}
\left< \psi^{2}(1)\psi^{2}(2) \right>_{n}
&=&
\left< 
\delta(\xi_{k}-x_{1})\delta(\xi_{l}-x_{2})
\right>^{cl}_{n}
\end{eqnarray}
with
\begin{eqnarray}
\left< {\cal O} \right>^{cl}_{n}
&\equiv&
\mbox{Tr}\left[ {\cal O} e^{-H_{n}}\right]/\mbox{Tr}\left[e^{-H_{n}}\right],
\nonumber \\
\mbox{Tr}
&\equiv&
\frac{1}{n!}\int \frac{d^{2}\xi_{1}}{a^{2}}\cdots \frac{d^{2}\xi_{n}}{a^{2}},
\nonumber \\
-H_{n}
&\equiv&
 4\sum_{ k<l }^{ n }{ \cal G }(\xi_{k},\xi_{l}),
\nonumber
\end{eqnarray}
where
$G(x_{i},x_{j}) \equiv \left< \Phi (x_{i}) \Phi (x_{j}) \right>$ 
is the Green's function
and ${\cal G} (x_{i},x_{j}) \equiv G(x_{i},x_{j})-G(0)\sim -g/2\pi\ln \left( |x_{ij}|/a \right)$.
Here, we multiply some trivial factors which reduce to unity in the limit $n\rightarrow 0$.
As is suggested in Eq. (\ref{stat}),
$\left< \psi^{2}(1)\psi^{2}(2) \right>_{n}$
can be interpreted as the two body density
of a 
{\em classical statistical system}
consisting of a set of particles (replicas)
interacting each other via the potential 
$
{\cal G}(x_{i},x_{j})
$\cite{rem2}.
These replica estimations are shown and 
directly compared to the numerical results (Fig. \ref{replica}).
In Fig. \ref{replica},
$\left<\psi^{2}(1)\psi^{2}(2) \right>_{n}$ for various 
$n$ (the number of replicas) with fixed $g$ are obtained 
by calculating Eq. (\ref{stat}) numerically on a $64 \times 64$ lattice.
This rather small lattice size is due to the multiple integral in Eq. (\ref{stat}).
Note that we do not have $\left<\psi^{2}(1)\psi^{2}(2) \right>_{n=1}$ 
as is inferred from Eq. (\ref{rep}).
For $g=0.4$, the replica estimation seems to converge 
to the one calculated by the direct numerical simulations.
For $g=6.4$, in contrast, 
it hardly seems to coincide to the exact one in the limit $n\rightarrow 0$.
Moreover, it gives an unphysical result, 
{\it i.e.}, a {\em negative} exponent, after taking the replica limit.

This breakdown is closely related to the transition {\em in the replica space}.
There are two distinct phases for this system.
For small $g$, 
all configurations are equally favorable.
As $g$ increases, however, the configurations 
where all replicas are close to each other come to have large weight.
Thus, for sufficiently small $g$,
it is enough to concentrate on only $\xi_{k}$ and $\xi_{l}$ in Eq. (\ref{stat}).
The configuration of the other particles is smeared out in the ensemble average
and irrelevant for $\left<\psi^{2}(1) \psi^{2}(2)\right>_{n}$.
For large $g$, on the other hand,
the main contributions in the ensemble average are from the configurations
where all replicas except $\xi_{l}$ (or $\xi_{k}$) 
are at $x_{1}$ ( or $x_{2}$, respectively).
Then, $\left< \psi^{2}(1)\psi^{2}(2)\right>_{n}$ is expected to behave as
\begin{eqnarray}
\left< \psi^{2}(1)\psi^{2}(2)\right>_{n}\sim1/|x_{12}|^{\Delta_{n}},
\nonumber
\end{eqnarray}
with
\begin{eqnarray}
\label{delta}
\Delta_{n}
\sim\cases{
2g/\pi  & for small $g$\cr
2g(n-1)/\pi  & for large $g$\cr
}.
\end{eqnarray}
Numerically calculated $\Delta_{n}$
is shown in Fig. \ref{exp},
which confirms the above estimation for small $n$ is reasonable. 
Taking the replica limit $n\rightarrow0$, we obtain $\Delta_{n=0}=2g/\pi$ for small $g$
which is consistent with the results obtained
by SUSY technique \cite{mudry}.
For the strong disorder regime, however,
the exponent reduces 
to $-2g/\pi$ which is unphysical.

Does it mean the replica trick is a mere trick ?
One of the possible scenarios is that 
the apparent breakdown is  
because we only take the small number of replicas into account.
So, for large $n$, the $n$ dependence of $\Delta_{n}$ 
may deviate from Eq. (\ref{delta}) and give the correct answer in the replica limit \cite{ryu}.
Moreover, the replica symmetry breaking (RSB) solution which was proposed for the free energy\cite{carpentier}
may be applicable also for the correlation function.

We also investigated other types of correlation functions 
such as $\left< \psi(1) \psi(2)\right>$ or 
$\left< \Phi(1)\psi^{2}(1)\Phi(2)\psi^{2}(2)\right>$
, the latter of which is of interest 
because it is related to the second derivative of the free energy 
$\left< \ln Z\right>/\ln \left( L/a \right)$
which shows the non-analyticity at $g=2\pi$.
Their behaviors are qualitatively similar to that of 
$\left<\psi^{2}(1)\psi^{2}(2)\right>$
in that, for small $g$, 
these correlation functions become steeper and steeper
as $g$ increases whose $g$ dependence are calculable by the replica trick.
For large $g$, however, their  $g$ dependencies are rather weak 
and the naive replica trick fails.

Another interesting approach is to utilize the formula 
$
\frac{1}{Z^{N}}=\frac{1}{(N-1)!}\int_{0}^{\infty} d\mu e^{-\mu Z}\mu^{N-1}
$
and express the correlation function as
\begin{eqnarray}
\label{liouville}
\left< \psi^{2}(1)\psi^{2}(2) \right> 
=\int_{0}^{\infty}d\mu \mu
\int {\cal D}\Phi e^{-2\Phi(x_{1})-2\Phi(x_{2})} e^{-S_{LFT}[\Phi]}
\end{eqnarray}
where 
$
S_{LFT}
=\int \frac{d^{2}\xi}{a^{2}}
\left[ 1/2g \left(\nabla\Phi\right)^{2}
+\mu e^{-2\Phi(\xi)}
\right]
$
\cite{tsvelik}.
This action resembles that of the Liouville field theory 
in two-dimensional quantum gravity.
However, since it was pointed out that there are some 
subtleties about the field theoretic treatment 
of  Eq. (\ref{liouville}) \cite{carpentier},
we evaluate it directly by using the replica estimates.
We expand $e^{-\mu Z}$ to express 
$\left<\psi^{2}(1)\psi^{2}(2)\right>$
as the superposition of the replica estimates 
with a different number of replicas,
{\it i.e.}, the grand canonical ensemble
\begin{eqnarray}
\left< \psi^{2}(1)\psi^{2}(2) \right> 
=
2\int_{-\infty}^{\infty}d\widetilde{\mu} 
\nonumber \\
\sum_{n=2}^{\infty}
(-1)^{n} e^{ -{\cal F}_{n}(\widetilde{\mu})}
\left<
\sum_{k\neq l}^{n} \delta(\xi_{k}-x_{1}) \delta(\xi_{l}-x_{2}) 
\right>_{n}^{cl}
\nonumber
\end{eqnarray}
where ${\cal F}_{n}(\widetilde{\mu})
=-\ln \mbox{Tr}e^{-( H_{n}-\widetilde{\mu}n)}+2n^{2}G(0)$
and $\mu=e^{\widetilde{\mu}}$.
Since for the weak disorder, 
the summand only trivially depends on $n$, 
we can easily sum up this suggestive expression 
and obtain the same result as the replica trick.

This method is applicable also for the free energy.
We expand 
$\ln Z = 
\int_{0}^{\infty} \frac{d\mu}{\mu} \left( e^{-\mu}-e^{-\mu Z}\right)$ as
\begin{eqnarray}
\left< \ln Z \right>
=
\int_{-\infty}^{\infty} 
d\widetilde{\mu} 
\left(
e^{-e^{ \widetilde{\mu} }}
-1
-\sum_{n=1}^{\infty}
(-1)^{n}e^{-{\cal F}_{n}(\widetilde{\mu})}
\right).
\nonumber
\end{eqnarray}
It is difficult to perform the summation 
for the strong disorder though  
we can obtain the correct answer for the weak disorder.
However, if we {\em simply} employ the RSB estimate
by Carpentier and Doussal\cite{carpentier},
$
e^{-{\cal F}_{n}}=n!^{-1}
\left( L/a \right)^{ ( pg/\pi +2/p+\widetilde{\mu})n}
$
where $p=1$ for the weak disorder 
and $\sqrt{2\pi/g}$ for the strong disorder,
the summation reproduces the exact result 
both for the weak and strong disorder regime
\cite{ryu}.

We thank Y. Morita for fruitful discussions.
S.R. is grateful to T. Oka for useful comments.
Y.H. was supported in part by a Grant-in-Aid 
from the Ministry of Education, Science, and Culture of Japan.
The computation in this work has been partly done 
at the YITP Computing Facility 
and at the Supercomputing Center, ISSP, University of Tokyo.


\begin{references}
\bibitem{rem2}
Note that ${\cal G}(x_{k},x_{l})<0$ for $x_{k}\neq x_{l}$ 
and ${\cal G}(x_{k},x_{k})=0$ by definition.
\bibitem{ryu2}
The RSB estimate for ${\cal F}_{n}$  
is originally proposed for $0\le n < 1$
and 
it does not agree with the numerical
replica estimate for ${\cal F}_{n=1,2,3,4,5}$.
However, the success of the simple employment 
of the RSB estimate 
is mysterious and 
seems to be related to the nature of the RSB.
Further exploration is interesting future issue.
\end{references}
\end{document}